\documentclass[epj]{svjour}
\usepackage{graphics}
\begin{document}
%
%\title{KEK result}
\title{Search for a kaonic nuclear state via $^4$He({\it{K}}$^-$, {\it{N}}) reaction at rest}
\author{
%First author\inst{1} \and Second author\inst{2}% etc
% \thanks is optional - remove next line if not needed
%\thanks{\emph{Present address:} Insert the address here if needed}%
  M.~Iwasaki\inst{1,2} \and 
 H.~Bhang\inst{3} \and 
  J.~Chiba\inst{4} \and 
  S.~Choi\inst{3} \and 
  Y.~Fukuda\inst{2} \and
  T.~Hanaki\inst{4} \and 
  R.~S.~Hayano\inst{5} \and 
  M.~Iio\inst{1} \and 
  T.~Ishikawa\inst{5} \and  
  S.~Ishimoto\inst{6} \and 
  T.~Ishiwatari\inst{7} \and
  K.~Itahashi\inst{1} \and 
  M.~Iwai\inst{6} \and
 P.~Kienle\inst{7,8} \and 
  J.~H.~Kim\inst{9}  \and 
  Y.~Matsuda\inst{1} \and 
  H.~Ohnishi\inst{1} \and 
  S.~Okada\inst{1} \and 
  H.~Outa\inst{1} \and 
  M.~Sato\inst{1,2} \and
  S.~Suzuki\inst{6} \and 
  T.~Suzuki\inst{1} \and
  D.~Tomono\inst{1} \and 
  E.~Widmann\inst{7} \and 
  T.~Yamazaki\inst{1,5} \and
  H.~Yim\inst{3}
}                     % Do not remove
%
%\offprints{}          % Insert a name or remove this line
%
\institute{
%Insert the first address here \and the second here
 \it Nishina Center for Accelerator-Based Science, RIKEN \and
 \it Department of Physics, Tokyo Institute of Technology \and
 \it School of Physics, Seoul National University \and
 \it Department of Physics, Tokyo University of Science \and
\it Department of Physics, University of Tokyo \and
\it IPNS, KEK (High Energy Accelerator Research Organization) \and
 \it Stefan Meyer Institut f\"ur Subatomare Physik \and
 \it Physik Department, Technische Universit\"at M\"unchen \and
 \it Korea Research Institute of Standard and Science 
}
\date{Received: date / Revised version: date}
% The correct dates will be entered by Springer

%
\abstract{
%Insert your abstract here.
Very recently, we have performed a couple of experiments, 
{\it{KEK PS-E549/E570}}, for the detailed study of the strange tribaryon $S^0(3115)$ 
obtained in {\it{KEK PS-E471}}. These experiments were performed to accumulate 
much higher statistics with improved experimental apparatus
especially for  the better proton spectroscopy of the $^4$He({\it{stopped K}}$^-$, {\it{N}}) reaction. 
In contrast to the previous proton spectrum, no narrow ($\sim$ 20 MeV) peak structure 
was found either in the inclusive $^4$He({\it{stopped K}}$^-$, {\it{p}}) 
or in the semi-inclusive $^4$He({\it{stopped K}}$^-$, {\it{p}}$X^\pm$) reaction channel, 
which is equivalent to the previous $E471$ event trigger condition. 
Detailed analysis of the present data and simulation shows that the peak, corresponding to 
$S^0(3115)$, has been an experimental artifact. 
Present analysis does not exclude 
the possible existence of a much wider structure. 
To be sensitive to such structure and 
for better understanding of the non-mesonic $K^-$ absorption reaction channel, 
detailed analysis of the data is in progress.
 % end of PACS codes
} %end of abstract
\maketitle
\section{Introduction}
\label{intro}
%Your text comes here. Separate text sections with
Recently, the {\it{KEK PS-E471}} group reported the observation of peak formation in their {\it{proton}} 
spectrum, 
which corresponds to the formation of a three
baryon system with strangeness $-1$ and isospin 1 \cite{Suz04}. 
This experimental search was motivated by the theoretical prediction by Akaishi and Yamazaki 
 \cite{PRC02}, that the $\overline{K}$ meson may form strong bound states 
 by several  light nuclei with an isospin zero. 
The most stable one is predicted to be formed by $K^-$ with $^3$He core, 
so that the experimental search aiming at $neutron$ spectroscopy was performed 
by using the $^4$He({\it{stopped K}}$^-$, {\it{n}}) reaction. 

   If one assumes that the peak corresponds to the kaonic nuclear bound state, then it means that 
the none-zero isospin state is formed with extremely large ($\sim$ 200 MeV) binding energy, 
while the predicted state has isospin zero with binding energy about 100 MeV. 
These are very outstanding features. If the assumption is true,  
the formed system could be more dense than the original theoretical prediction.  
Therefore, experimental confirmation is strongly required. 

\section{Upgraded Experimental Setup}
\label{setup}
\begin{figure}
	\resizebox{0.48\textwidth}{!}{%
	  	\includegraphics{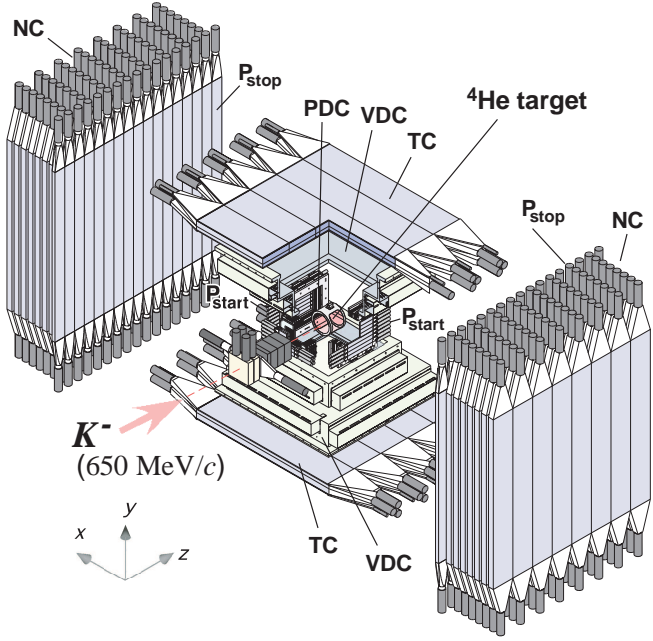}
	}
	\caption{Present experimental setup. 
	The setup is constructed symetrically around the central liquid helium target.
	Left and right arm are neutron counter arrays (NC). In front of NC, new segmented proton 
	TOF stop counters were installed. These horizontal counters are used 
	to achieve the proton inclusive spectroscopy.
	Top and bottom counters are primarily for the 
	decay charged particle $X^\pm$ detection, which is 
	 indispensable for the neutron spectroscopy or the detailed analysis using 
	 decay particles. 
	}
	\label{experimental_setup}       % Give a unique label
\end{figure}

\begin{figure}
	\resizebox{0.48\textwidth}{!}{%
	  	\includegraphics{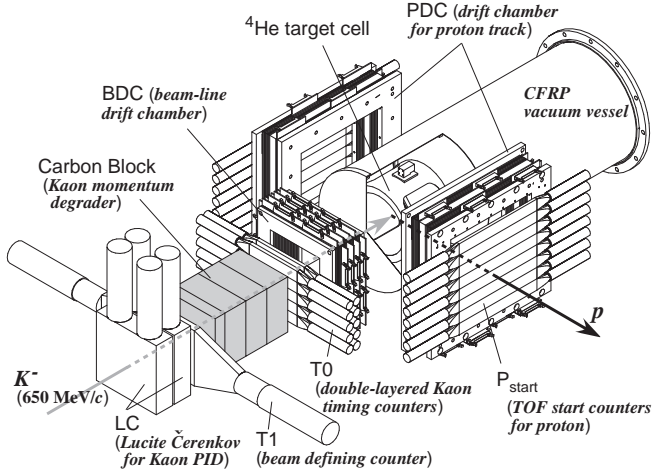}
	}
	\caption{Close up view of the  experimental setup around the target region. 
	Segmented proton TOF start counter arrays are placed behind the proton 
	tracking chambers (PDC). The segmented neutron TOF start counters (T0), 
	placed in front of the beam kaon tracking chamber (BDC), were upgraded to 
	the double layered structure to achieve better resolution of the neutron TOF analysis. 
	}
	\label{experimental_setup-target}       % Give a unique label
\end{figure}

We improved the experimental setup as shown in 
Figs. \ref{experimental_setup} and  \ref{experimental_setup-target}, 
based on the $E471$ setup (given in reference  \cite{Suz04}), 
to achieve confirmatory experiment shortly before KEK-PS shutdown. 
  The previous $E471$ setup was optimized to perform neutron spectroscopy 
by the time-of-flight (TOF) method 
from the ({\it{stopped K}}$^-$, {\it{n}}$X^\pm$) reaction on a liquid  $^4$He target, 
where $X^\pm$ is one of the decay charged particles detected by the top and bottom trigger counter system (TC). 
  The neutron TOF can be calculated from the time difference between 
incoming kaon timing at T0 counters and one of the NC array, by subtracting 
kaon stopping time using its range information calculated from the vertex 
between $K^-$ and $X^\pm$ tracks. 

  The energy resolution and detection efficiency 
  for neutrons was improved
by replacing single-layered segmented kaon timing counters 
T0 with double-layered ones, and 
enlarging the number of neutron counter arrays (NC). 
  For the neutron spectroscopy,  the track information of this additional charged 
particle $X^\pm$ on TC is indispensable even with the present setup
to identify the kaon reaction point.

  On the other hand, proton TOF can be performed more directly with dedicated start and stop counters and the proton track information itself. 
 Therefore, we newly installed drift chambers (PDC) and fine segmented start- (P$_{\rm{start}}$) and stop-
(P$_{\rm{stop}}$) counters for proton spectroscopy  
 as shown in Fig. \ref{experimental_setup-target}.
 We removed thin charged-veto counters and iron plates in front of the NC array, instead, to 
 enlarge the acceptance for low momentum protons. 
  With this setup, we can perform inclusive proton spectroscopy of the 
$^4$He({\it{stopped K}}$^-$, {\it{p}}) reaction 
over a wide momentum range 
without requiring an additional $X^\pm$ track on TC. 
  The statistics of the inclusive proton spectrum can be significantly improved compared to the 
previous semi-inclusive proton spectrum of the $^4$He({\it{stopped K}}$^-$, {\it{p}}$X^\pm$) reaction, 
whose statistics were limited by the solid angle of TC.
  In the present analysis, we can also check the 
  consistency of our momentum analysis procedure 
using redundant information of proton TOF between T0, 
p$_{\rm{start}}$ and p$_{\rm{stop}}$ counters.

We also improved particle identification (PID) capability of the TC counters for $X^\pm$
by adding additional two layers to the previous $E471$ setup (namely 2+2 layers), 
because high momentum protons above $\sim$ 500 MeV/$c$ cannot be
separated from pions in the $E471$.
The PID analysis on TC shows that we can discriminate 
pions from protons over a wide
momentum range from 300 $\sim$ 600 MeV/$c$, 
so that we can reconstruct the hyperon invariant mass using $\pi N$-pairs detected in TC and NC.

   We performed two experiments $E549/E570$ at KEK 
both aimed at confirmation   
of the 
S$^0$(3115) in the improved proton spectrum, 
and obtaining higher neutron statistics for a detailed study
together with the other decay products.
   $E549$ is the dedicated experiment for $^4$He({\it{stopped K}}$^-$, {\it{N}})
spectroscopy.  
We also accumulated data parasitically in E570, 
whose primary object is the precise measurement of the $3d$ $\rightarrow$ $2p$
x-ray energy  in kaonic helium atoms.			
   By using roughly seven times the beam time compared to $E471$, we accumulated 
about 50 times more statistics for proton inclusive data and 10 times more for neutron.
   In the case of neutron analysis, 
we do need $X^\pm$ recorded on TC as in the previous experiment, 
so that the statistics improvement comes mainly from the larger NC counter 
volume. 

After the slewing correction, to compensate the time-walk caused by the finite discriminator threshold, 
the TOF resolution (RMS) was improved from 300 $\rightarrow$ 120 psec for protons and 
300 $\rightarrow$ 200 psec for neutrons. 
The difference of time resolution improvements comes from the 
TOF definition between the two.
For proton TOF between P$_{\rm{start}}$ and P$_{\rm{stop}}$, 
the slewing function is calibrated by using $K\mu$II and $K\pi$II decay 
events from $K^+$ runs.
%For neutron TOF between T0 and one of NC array, 
%it is calibrated by using $\gamma$-ray events of $K^-$ run.
Neutron TOF is measured between T0 and the NC array, instead, 
and it is calibrated using $\gamma$-ray events in $K^-$  runs.

\section{Proton spectra}

The primary object of the upgraded experiment at KEK is to confirm whether or not the 
previous interpretation of the $S^0$(3115) peak formation in the proton spectrum is true.
Let us focus on the proton spectroscopy in this paper, because  
the neutron spectroscopy will be given in another paper \cite{Yim07}.

\begin{figure}
	\resizebox{0.45\textwidth}{!}{%
	  	\includegraphics{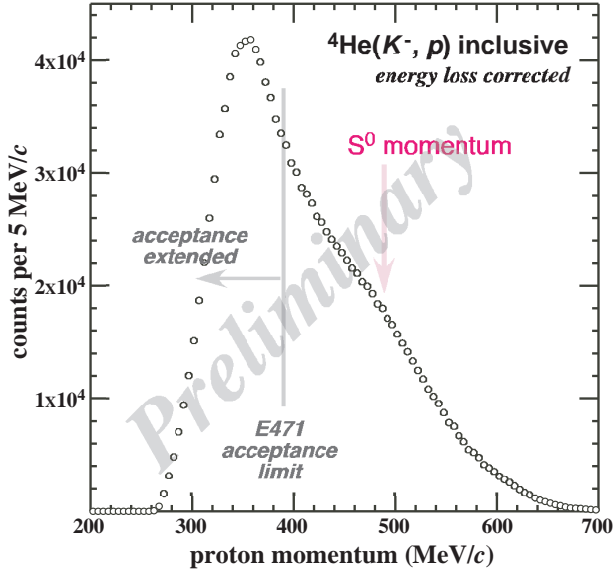}
	}
	\caption{Inclusive proton momentum spectrum. The proton TOF is measured 
	by the time difference between P$_{\rm{start}}$ and P$_{\rm{stop}}$ counters. 
%	The original proton momentum at the reaction vertex (horizontal axis) is evaluated iteratively 
%	using energy-loss calculation code to have consistent TOF data, 
%	taking into account the energy loss before P$_{\rm{stop}}$ counter, 
%	namely target, P$_{\rm{start}}$ counter, and the air in-between.
	}
	\label{inc-p-mom}       % Give a unique label
\end{figure}

The inclusive $^4$He({\it{stopped K}}$^-$, {\it{p}}) proton TOF spectrum between  
P$_{\rm{start}}$ and P$_{\rm {stop}}$  counters with PDC tracking information
is shown in Fig. \ref{inc-p-mom}. 
The proton PID on NC counters is performed by using the correlation between 
$1/\beta$ and total energy deposit in the P$_{\rm{ stop}}$ and the NC array.
The initial proton momentum was computed iteratively using energy-loss calculation code, 
so as to obtain consistent P$_{\rm{start}}$ and P$_{\rm{stop}}$ timing by assuming 
that the proton is originated from the primary kaon reaction point given by the vertex 
determined from kaon and proton tracks. Because of target, P$_{\rm{start}}$, and the air, 
the momentum differed from simple TOF calculation by about 30 MeV/$c$. 
The error of the momentum caused by the analysis procedure is small enogh compared to 
the time resolution, except for the proton caused by the hyperon decay.
%In contrast to the previous $E471$ experiment,
%the peak in proton spectrum corresponding to 
%the S$^0$(3115) formation is not seen, in spite of the improved resolution. 

The spectrum is much different  
from that of the previous $E471$ experiment \cite{Suz04},
i) the statistics are drastically improved,  
ii) the proton momentum acceptance is extended to the low momentum side substantially, 
and iii) there is no clear peak structure at the momentum where we expected the signal from 
$S^0$(3115) formation.

Why does the signal disappear in
the inclusive proton spectrum?
It is extremely important to examine the 
$^4$He({\it{stopped K}}$^-$, {\it{p}}$X^\pm$) spectra. 
%The most simple explanation of the present spectrum would be that the 
%$S^0$(3115) signal in $E471$ data could be formed or enhanced due to the previous 
%trigger condition, in which the $X^\pm$ detection on TC is requested in the hardware level. 
%It should also be noted that t
Most significantly, the peak was seen clearly in the charged-pion-tagged spectrum, but
%The biggest reason, why the reference \cite{Suz04} excluded 
%the possibility of the experimental artifact, is that the peak is seen clearly in the 
%charged pion tagged spectrum, while 
it was not in the proton tagged one in $E471$.
If the $S^0$(3115) peak formation was due to the simple experimental defect, 
it would be seen in both 
spectra. 

\begin{figure}
	\resizebox{0.45\textwidth}{!}{%
	  	\includegraphics{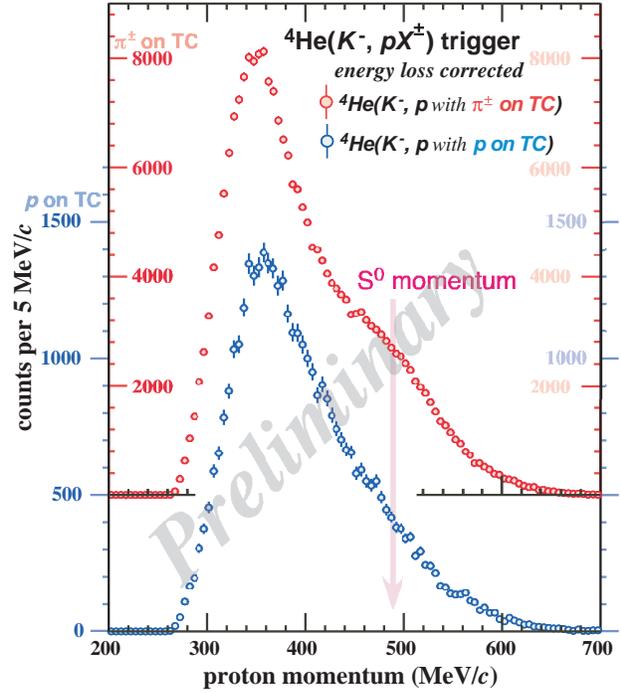}
	}
	\caption{Proton momentum spectra of $^4$He({\it{stopped K}}$^-$, {\it{p}}$X^\pm$) 
	reaction. Spectra of charged pion (right-scale) and proton (left-scale) tagged on TC 
	are shown separately. 
	These spectra are the subset of Fig. 3.
	Note that the 
	vertical scales are different between the two.
	}
	\label{tagged-mom}       % Give a unique label
\end{figure}

Figure \ref{tagged-mom} shows the $^4$He({\it{stopped K}}$^-$, {\it{p}}$X^\pm$) spectra. 
The statistics of both are substantially reduced 
due to the solid angle of TC. 
The decay particles $X^\pm$ on TC are required in the software level 
to reproduce the $E471$ trigger condition, 
although peak formation is seen in
%though the peak formation are seen in 
neither of the proton spectra. 
Therefore, the result 
is not consistent with $E471$. 
%is inconsistent to that of the $E471$ one.

Note also the spectral difference between the two at momentum over 
400 MeV/$c$.  
A rounded elevation is seen in the charged-pion 
tagged spectrum, while a monotonic decrease results from proton tags. 
Before discussing the spectral difference, let us focus on the 
reason  why a narrow peak, $S^0$(3115), was seen in the $E471$  
proton spectra.

\section{Difference of the proton TOF analysis}
It is crucial to understand why and how the results are inconsistent to each other.
The present data indicates that some of the previous analysis procedure might give a
spectral-singularity slightly below 500 MeV/$c$ in the proton spectra.
%, could be hidden 
%and unresolved in the previous $E471$ proton TOF analysis. 

Actually, this is the momentum where a proton stops at the back side of the proton 
TOF stop counter in $E471$ (NC first layer in that setup). 
Therefore, we performed Monte Carlo simulations to study the light-output 
response of the NC first layer as a function of the original proton momentum at the 
reaction point 
assuming the configuration of the
previous $E471$ experimental setup.

\begin{figure}
  \begin{center}
	\resizebox{0.4\textwidth}{!}{%
	  	\includegraphics{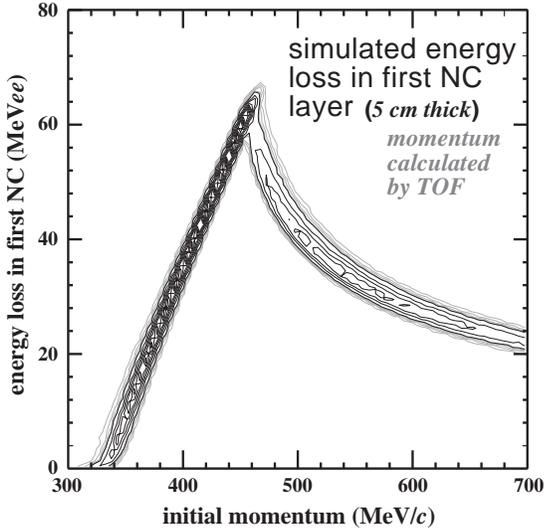}
	}
  \end{center}
	\caption{Contour plot of the simulated energy loss in the proton TOF stop counter (NC first layer 
	as for the $E471$ experimental configuration whose thickness is 5 cm).
	The light saturation at high $dE/dx$  (Birk's effect) is taken into account. 
	}
	\label{sim-mom-dE}       % Give a unique label
\end{figure}

Figure \ref{sim-mom-dE} is the result of the simulation, where the horizontal axis 
is the proton initial momentum and the vertical axis is the expected light output from 
the plastic scintillator with the thickness of 5 cm. 
In the simulation, reaction point distribution and the counter resolution 
(light collection efficiency to the PMTs at both ends) are taken into account, which makes the distribution band width to be wider.

The light output distribution in the counter has cusp-like structure at around 460 MeV/$c$. 
The lower momentum side gives less light output, 
because the range is shorter than the counter thickness. 
The protons whose momentum below 320 MeV/$c$ never reach the TOF stop counter.
The light 
gets smaller again on the higher momentum side as they approach minimum ionizing
 particles (MIPS) $\sim$ 10 MeV$ee$ (electron equivalent) in 5 cm thick 
plastic counter. The sudden drop of the light output is due to the nature of the Bragg curve, 
which has maximum energy loss before the charged particle stops.

This cusp-like singularity may cause a severe problem in the TOF analysis procedure.  
As it is described, TOF start and stop counters hit-time information is corrected so as to have  
best resolution for MIPS ($K\mu$II events). 
The typical slewing correction function is written as:
\begin{eqnarray}
\label{t_correction_0}
	t =\frac{T_A + T_B}{2} - (\frac{c_A}{\sqrt{A_A}} + \frac{c_B}{\sqrt{A_B}}) - T_0,
\end{eqnarray}
where $T_A$ and $T_B$ are the TDC data of the PMT at both ends of a counter, 
$A_A$ and $A_B$ are that for charge sensitive ADC for the signal pulse, 
$c_A$ and $c_B$ are the correction parameters, and $T_0$ is time-offset of the counter. 
The parameters $c_A$, $c_B$ and $T_0$ were defined and calibrated for each counter 
to minimize the $1/\beta$ $(= c\Delta t / \Delta L)$ distribution width at 
known $1/\beta$ value for the MIPS. 

The problem is that the correction formula (\ref{t_correction_0}) depends on 
the energy deposit in the counter. 
The typical energy deposit in the cusp region is $\sim$ 6 times higher in energy 
than that of MIPS (calibration point) so that the 
simple application of the formula may 
result in a
deformation of the TOF spectrum 
at around 460 MeV/$c$.

\begin{figure}
  \begin{center}
	\resizebox{0.4\textwidth}{!}{%
	  	\includegraphics{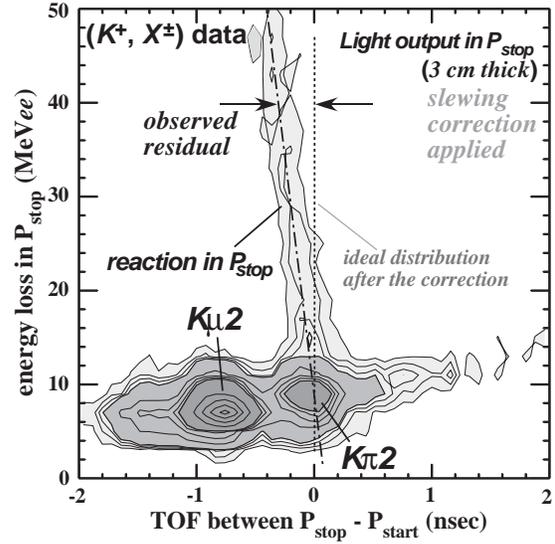}
	}
  \end{center}
	\caption{Energy loss observed in P$_{\rm{stop}}$ (3 cm thick) counter for $(K^+, X^\pm)$ events
	obtained in $E549$. Simple slewing correction defined by MIPS has been applied.
	The horizontal axis is the TOF between P$_{\rm{start}}$ and P$_{\rm{stop}}$, and the vertical axis 
	is the light output of P$_{\rm{stop}}$ counter.
	}
	\label{sato_dE}       % Give a unique label
\end{figure}

Figure \ref{sato_dE} is a contour plot of $1/\beta$ and energy deposit in the TOF stop 
counter (P$_{\rm{stop}}$) of $E549$. 
The horizontal axis is the $1/\beta$ obtained from P$_{\rm{start}}$ and P$_{\rm{stop}}$ 
counters after the simple slewing correction (\ref{t_correction_0}) has been 
applied. The vertical axis is the position-averaged light output of the P$_{stop}$ counter in the form of 
$\sqrt{A_A A_B}$ ($=\bar{A}$) to cancel the light attenuation effect in the first order.
In the figure, a higher energy component is seen 
originating from the $K\pi$II peak. 
The light output of these events in the P$_{\rm{start}}$ counter is consistent with MIPS, so that the 
large energy observed in the P$_{\rm{stop}}$  counter should be the result of nuclear reaction 
of the pion in P$_{\rm{stop}}$.

\begin{figure*}
% Use the relevant command for your figure-insertion program
% to insert the figure file. See example above.
% If not, use
\begin{center}
	\resizebox{0.76\textwidth}{!}{%
	  	\includegraphics{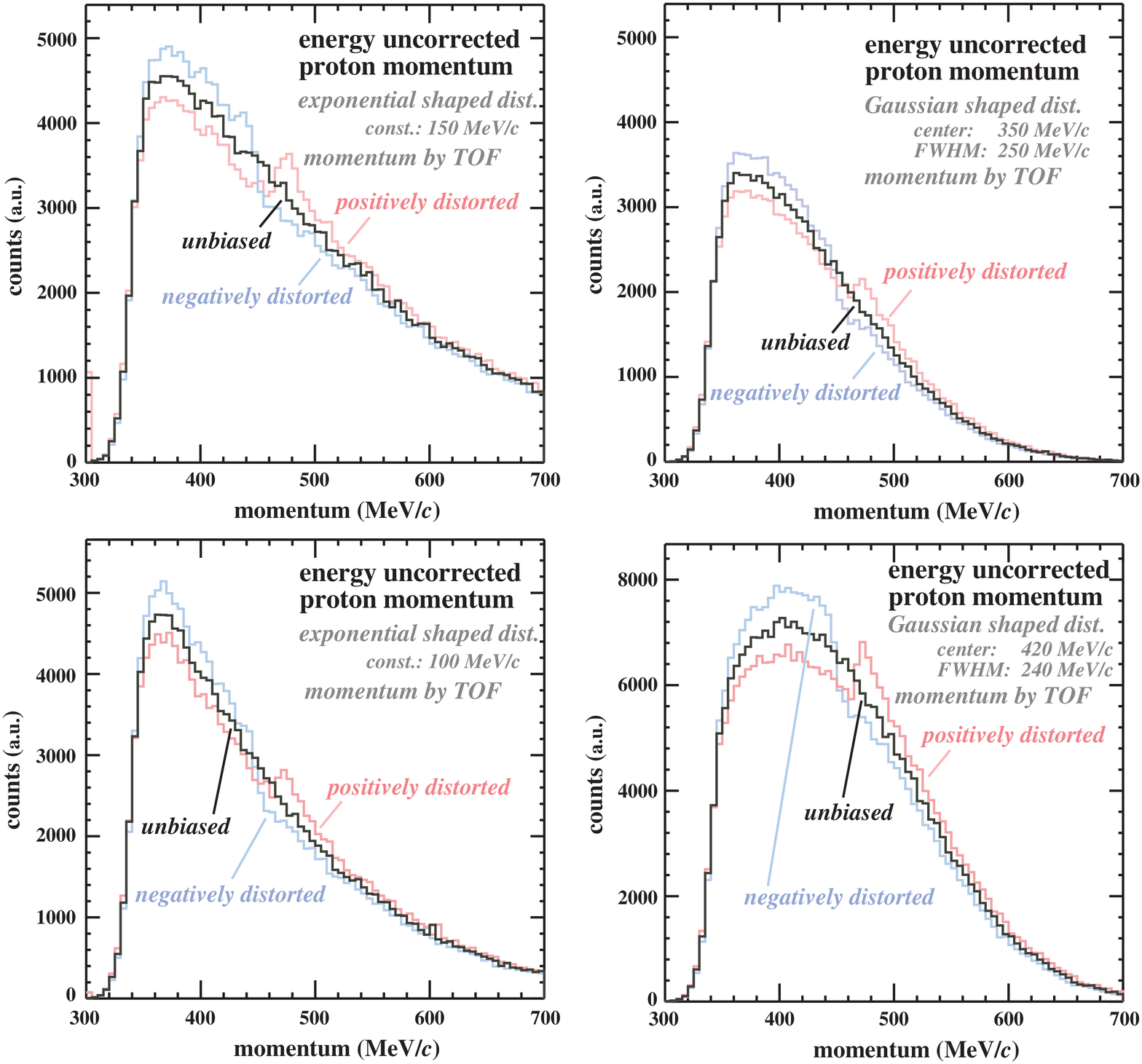}
	}
\end{center}
%\vspace*{5cm}       % Give the correct figure height in cm
\caption{Simulated distortion effect due to the residual time versus pulse-hight of the proton TOF spectra.
Upper left and lower left panels are the result of exponential proton spectra with 
the constant of 150 and 100 MeV/c, respectively. Upper right and lower right ones are 
for Gaussian distribution center and width of 350 and 250 (upper), and 420 and 240 MeV/c 
(lower), respectively. TOF linear distortion of   $\pm$ 5 $psec$/MeV$ee$ is assumed 
for all the spectra. The momentum loss of the proton is not corrected.}
\label{fig:2}       % Give a unique label
\end{figure*}

If the slewing correction function (\ref{t_correction_0}) 
%which is calibrated by MIPS, 
%is precise enough to apply 
is applicable 
to the large energy deposit region, 
no correlation is expected in this figure. 
However, the pion reaction events are inclined towards 
upper-left of this figure.
It clearly indicates that the slewing correction by the formula  (\ref{t_correction_0}) 
is not sufficient, and most of the counters have similar residuals as a function of energy 
deposit in the counter. 
%there exist time and pulse-hight correlation 
%between the TOF determination logic and 
%the deposit energy at the TOF stop counter. 
In the previous $E471$ experiment, the TOF calibration was performed using  
$\gamma$-ray events. The $\gamma$-ray was converted to the electron shower at the 
iron plate located in between the  NC array and the thin charged-veto counters 
(removed from the present setup). 
Very unfortunately, the $\gamma$-ray originated signals don't give large energy deposit in 
the TOF stop counter.
Therefore, there is no good calibration data at the high energy deposit region. 
%in the previous $E471$ experiment.
 %tail toward 
%higher energy region.
%This prevent to realize the hidden time and pulse-hight correlation in the previous analysis. 

The simplest term to cancel the residual correlation between time and pulse-hight can be written as:
\begin{eqnarray}
\label{t_correction_1}
	\Delta t = c_R \left(\bar{A} - A_0 \right),
\end{eqnarray}
where $A_0$ is the average ADC count for MIPS.
In the case of $E549/E570$ data, a more precise correlation function could be defined for
each counter, though the formula  (\ref{t_correction_1})
is enough to simulate what may happen in the previous $E471$ analysis.

The effect of the hidden-residual term in the proton TOF analysis 
to the proton momentum spectrum can be reproduced 
by intentionally adding the residual to the simulated one. 
We have performed the simulations by assuming several artificial proton 
spectra with the deformation parameter $c_R$ as small as $\pm$ 5 $psec$/MeV$ee$. 

As shown in the figures, all the spectra have distortion 
slightly above the cusp region where $S^0$(3115) was found in $E471$ analysis, 
when the TOF is distorted in the positive direction.
The peak structure is more clearly seen when the original proton spectrum 
has rounded elevation around the momentum (lower-right pannel of the figures). 
This is consistent with the result of $E471$ analysis that the $S^0$ is seen more 
clearly in the pion tagged distribution, which have rounded elevation around the momentum.

Therefore, it is concluded that the $S^0$(3115) peak observed in 
proton spectrum in E471 is most likely formed as an experimental artifact, 
when the range of the particle matches to the counter thickness.
On the other hand, such artificial peak formation cannot be expected in the neutron spectrum, 
because the neutron can be detected at any depth of the NC counter.
%It should also be noted that the artifact is originated the nature of the Bragg's curve. 
%Therefore, such artificial peak formation cannot be expected in the neutron spectrum.

\section{Upper limit of the peak formation}

In the present experiment, one of the important questions to be answered is what is 
the upper limit of formation of the kaonic nuclear bound state from the inclusive 
proton spectrum. 

To obtain the upper limit of the formation of kaonic nucleus per kaon reaction at rest, 
we need to evaluate 
the number of the stopped kaon in the helium target and 
proton acceptance as a function of its momentum. 
We evaluated the number using the known free decay branch of {\it{ so called}} 
meta-stable kaons 3.5 $\pm$ 0.5 \% \cite{OTA} (a negative 
kaon in the atomic orbit of helium at a large angular
momentum).
The proton detection efficiency has momentum dependence, because of 
the range at the low momentum side and the proton PID at the high momentum side.   
We identified the proton using its momentum and total energy recorded in the NC array. 
The higher momentum side of the proton spectrum given in Fig. \ref{inc-p-mom} has
less efficiency compared to the low momentum side, due to the reduction of the effective solid angle 
and reaction loss of the proton. 
%Therefore, higher missing mass region of Fig. \ref{inc-p-missing-mass} 
%is more stressed than the lower side. 
The proton detection efficiency is calculated by a simulation.

\begin{figure}
	\resizebox{0.43\textwidth}{!}{%
	  	\includegraphics{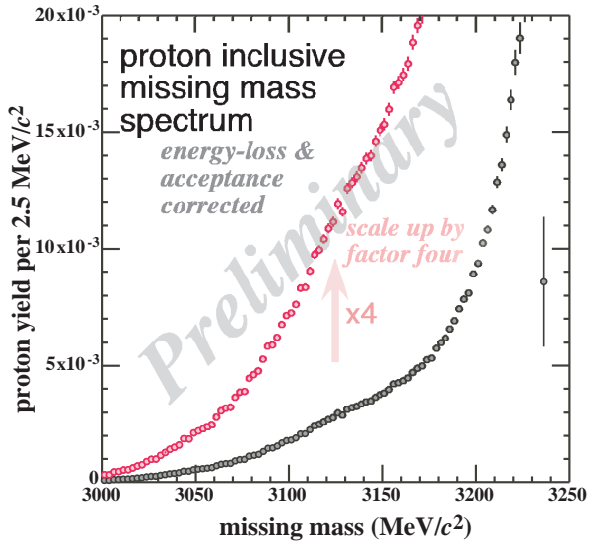}
	}
	\caption{Acceptance corrected proton missing mass spectrum. 
	}
	\label{inc-p-missing-mass}       % Give a unique label
\end{figure}

%\begin{figure}
%	\resizebox{0.45\textwidth}{!}{%
%	  	\includegraphics{upper-limit.eps}
%	}
%	\caption{Upper limit of the formation probability per stopped kaon reaction. 
%	Acceptance corrected proton spectrum was fitted  by third order polynomial function
%	with a Gaussian centered at the fit region. The upper limit is examined by Bayesian 
%	method with the Gaussian width ($\sigma$) of 3.5 10 and 20 MeV/$c^2$. 
%	}
%	\label{upper-limit}       % Give a unique label
%\end{figure}

Figure \ref{inc-p-missing-mass} shows the missing mass spectrum of protons 
after the acceptance correction. 
In this figure, the vertical axis is 
normalized by the number of stopped kaons. 
We have evaluated the upper limit 
by assuming a smooth proton spectral function (third polynomial function)
together with a Gaussian centered at 3115 MeV/$c^2$.
%, because there is no reliable proton spectral function. 
%We examined the upper limit by Bayesian 
%method with the Gaussian centered at 
%the fit region with the
%width ($\sigma$) of 3.5 10 and 20
%around 3115 MeV/$c^2$. 
%To check the stability of the 
%obtained upper limit, we fitted the spectrum with several mass window as indicated 
%in the figure and hatched according to the examined width, 
%except for the upper limit for $\sigma = 20$ MeV/$c^2$.
%
%As shown in he Fig. \ref{upper-limit}, the existence of narrow ($\sigma < $ 10 MeV/$c^2$) 
The upper limit of the narrow ($\Gamma < $ 20 MeV/$c^2$) peak
formation at this mass region is obtained to be 
%peak formation is severely excluded 
well below 10$^{-3}$ at the 95~\% confidence level.
This upper limit is quite severe compared to the at-rest kaon-induced 
hypernuclear formation probability of the order of \%. 
Therefore, the existence of 
such a narrow state
is not very likely. 

\section{Discussion and Conclusion}

As it is described, the signal of S$^0$(3115) observed in $E471$ is most likely due to the 
experimental artifact. 
This experimental problem exist only in proton, 
but not neutron spectroscopy.
The upper limit of the formation probability of such a narrow peak ($\Gamma < $ 20 MeV/$c^2$) 
is in the order of 10$^{-4}$ at S$^0$ energy.
However, it does not exclude the existence of a wider structure.

The sensitivity for the wider structure is limited because of the large background, so that it is 
very important to understand background components of the proton spectrum.

The dominant reaction branch of the kaon reaction at rest is known to be 
quasi-free hyperon production with pion emission, namely
$K^-N \rightarrow Y\pi$. 
If this is the primary reaction, then most of the energy is carried out by the pion kinetic energy, 
so that the proton (or nucleon in general) from the hyperon decay 
is dominantly produced at momenta lower than  
$\sim$ 400 MeV/$c$ ({\it cf.} Fig. \ref{tagged-mom}) 
or above $\sim$ 3200 MeV/$c^2$ in the missing mass spectrum ({\it cf.} Fig. \ref{inc-p-missing-mass}). 
The nucleons produced by $\Sigma$-$\Lambda$ conversion, 
$\Sigma N \rightarrow \Lambda N$,  are also located mostly in the same region.
%In this region, there would be all the other cascade reaction products.
%If the signal locates in this region, 
It is difficult to improve sensitivity to detect wider structures in this region, because 
the spectrum changes drastically depending on the momentum.

On the other hand, the higher momentum region above $\sim$ 400 MeV/$c$ is 
more simple. 
This region originates dominantly from non-mesonic kaon absorption,
$K^-NN \rightarrow YN$. 
Actually, 
the spectral difference between charged pion and proton tagged spectra shown in Fig.  
\ref{tagged-mom} is due to the sensitivity to the non-mesonic process. 
This is because of the kinematics of the hyperon decay, $Y \rightarrow N \pi$. 
The direction of $Y$ and $N$ is almost the same in the Lab. frame while the 
$\pi$ distribute almost uniformly, because of the small pion mass. 
Therefore, the pion tagged one is more sensitive to the non-mesonic kaon absorption
process, which gives rounded elevation of the spectrum at high momentum region.
%If the signal of the kaonic nuclear-state formation located in the high momentum region, 
%one may have better sensitivity by the study of hyperon motion. 
%If the tagged pion is from a hyperon decay, then the track can shift from the 
%kaon stop point due to the hyperon motion before the decay. 
 
%The study of spectral difference using the 
%hyperon motion was discussed in reference \site{Iwa04}.  
In the $E471$ analysis, improvement of the signal-to-noise ratio 
in this momentum region was intended 
using pion-trajectory defined hyperon motion \cite{Iwa04}. 
The original idea of the analysis is to try to improve the signal fraction 
by selecting low momentum hyperon in the final state. 
The $Q$-value of the background $K^-NN \rightarrow YN$ (non-mesonic) process is very large, 
so that the kinetic energy of the hyperon is expected to be larger than that of signal.
Actually, same analysis as reference \cite{Iwa04} gives 
similar spectral enhancement of broad structure in neutron spectra, 
but we need to study the data more carefully, 
because the kinematical difference between the signal and non-mesonic
is not extremely large, and the difference becomes more marginal 
if final state interaction of the hypeon is taken into account.

%Both signal and background processes have a hyperon in the final state.
%The major background process is the direct $K^-NN \rightarrow YN$ process.
%
%could be found 
%the kinetic energy of the hyperon 
%produced via kaonic nucleus should be lower than that of non-mesonic process. 
%It is not very straight forward, because the kinematics of the signal is 
%quite close to that of the non-mesonic kaon absorption. 
%Both processes have a hyperon in the final state, and 
%the difference between the two process is whether the hyperon is produced via 
%kaonic nucleus or direct $K^-NN \rightarrow YN$ process.
%Although, kinetic energy of the hyperon 
%produced via kaonic nucleus should be lower than that of non-mesonic process. 
%so that it is important to study the data more carefully.
%We should also take into account the final state interaction.
%
%Therefore, detailed study of the hyperon motion is very important. 
%In the $E471$ analysis, we intended to improve the signal-to-noise ratio  
%using pion trajectory defined hyperon motion \cite{Iwa04}. 
With the present higher statistics data, we are presently analyzing data 
based on the invariant mass similar to the FINUDA paper  \cite{Fuj05}. 
The analysis is in progress and 
the result will be reported in the near future.


\begin{thebibliography}{}
\bibitem{Suz04} T.~Suzuki et al., Phys. Lett. B \textbf{597},  (2004) 263. 
\bibitem{PRC02} Y.~Akaishi and T.~Yamazaki, Phys. Rev. \textbf{C 65}, (2002) 044005.
\bibitem{Yim07} H.~Yim et al., Proceedings of HYP06,   Johannes Gutenberg University Mainz. 
\bibitem{OTA} T. Yamazaki et al., Phys. Rev. Lett. \textbf{63}, (1989) 1590, \\
H. Outa Doctoral Dissertaion, Univ. of Tokyo (2003).
\bibitem{Iwa04} M.~Iwasaki et al., arXiv: nucl-ex/0310018. 
\bibitem{Fuj05} M. Agnello et al., Phys. Rev. Lett. \textbf{94}, (2005) 212303. 
\end{thebibliography}
\end{document}